\documentclass[aps,prb,twocolumn,longbibliography,showpacs]{revtex4-1}
\usepackage{ifthen}

\usepackage[pdftex]{graphicx} 
\usepackage{epstopdf}

\usepackage{amssymb,amsmath,amsfonts,hyperref}
\usepackage{wasysym}
\usepackage{latexsym}

\begin{document}
\title{Dirac's ``magnetic monopole'' in pyrochlore ice $U(1)$ spin liquids: Spectrum and classification}
\author{Gang Chen$^{1,2}$}
\email{gangchen.physics@gmail.com}
\affiliation{$^{1}$Department of Physics, 
Center for Field Theory and Particle Physics, 
State Key Laboratory of Surface Physics,
Fudan University, Shanghai 200433, China}
\affiliation{$^{2}$Collaborative Innovation Center of Advanced Microstructures,
Nanjing, 210093, China}

\date{\today}

\begin{abstract}
We study the U(1) quantum spin liquid on the pyrochlore spin ice systems.
For the non-Kramers doublets such as Pr$^{3+}$ and Tb$^{3+}$, we point out 
that the inelastic neutron scattering result not only detects the low-energy gauge 
photon, but also contains the continuum of the ``magnetic monopole'' 
excitations. Unlike the spinons, these ``magnetic monopoles'' are purely 
of quantum origin and have no classical analogue. We further point out 
that the ``magnetic monopole'' experiences a background dual ``$\pi$'' flux 
due to the spin-1/2 nature of the local moment when the ``monopole'' hops    
on the dual diamond lattice. We then predict that the ``monopole'' continuum 
has an enhanced spectral periodicity with a folded Brillouin zone. 
This prediction can be examined among the existing data on 
the non-Kramers doublet spin liquid candidate materials like 
Pr$_2$TM$_2$O$_7$ and Tb$_2$TM$_2$O$_7$ (with TM = ``transition metal''). 
The application to the Kramers doublet systems and numerical simulation is 
further discussed. Finally, we present a general classification of distinct
symmetry enriched U(1) quantum spin liquids based on the translation symmetry 
fractionalization patterns of ``monopoles'' and ``spinons''. 
\end{abstract}

\maketitle

\section{Introduction}
\label{sec1}

There has been a tremendous activity in the field 
of pyrochlore ice materials~\cite{PhysRevLett.98.157204,Gingras2014,BalentsSavary,PhysRevLett.105.047201,Savary12,Sungbin2012,SavaryPRB,PhysRevLett.87.067203,PhysRevB.65.054410,PhysRevLett.87.047205,Gingras2014,Ross2009,Huang2014,PhysRevB.94.205107,PhysRevLett.108.247210,PhysRevB.95.041106,PhysRevB.95.094422,Chen2015,SavaryPRB,PhysRevLett.109.017201,Yasui2002,PhysRevB.64.224416,PhysRevB.90.214430,Chang2012,Kimura2012,RevModPhys.82.53,Lhotel2014,Chang2014,Yasui2003,Ross11,Shannon12,Goswami2016,PhysRevB.95.094407,PhysRevLett.118.107206,PhysRevB.92.054432,PhysRevLett.113.197202,fu2017fingerprints,PhysRevB.86.075154,PhysRevLett.115.267208,PhysRevLett.109.097205,PhysRevLett.107.207207,PhysRevLett.115.097202,Shannon2017,GangChen2017,PhysRevLett.118.087203}. 
Besides the early efforts in classical spin ice and dipolar spin ice where 
quantum effects are negligible~\cite{Gingras2001,PhysRevLett.87.067203,Castelnovo12008}, 
a recent motivation of this exciting area is to 
search for the three-dimensional U(1) quantum spin liquid (QSL)~\cite{Hermele04} 
in the pyrochlore quantum spin ice systems where quantum effects 
are significant~\cite{PhysRevLett.98.157204,Savary12,Sungbin2012,PhysRevLett.105.047201}. 
The existence of this exotic quantum phase of 
matter has been firmly established by the theoretical studies of the relevant 
and even realistic spin models on the pyrochlore lattice~\cite{Hermele04,Savary12,Shannon12,Sungbin2012,Huang2014,Gingras2014,BalentsSavary,
PhysRevLett.100.047208,PhysRevLett.115.077202,PhysRevLett.115.037202}. 
The experimental confirmation of this interesting
phase of matter, however, is still open. Even if this phase may have already 
existed in some candidate materials since the original proposal
in Tb$_2$Ti$_2$O$_7$~\cite{PhysRevLett.98.157204} and Yb$_2$Ti$_2$O$_7$~\cite{Ross11,Savary12}, 
the firm identification of this exotic phase requires the strong mutual feedback between 
the experimental progress and the theoretical development 
that provides and clarifies unique and clear physical observables for the
experiments. 

The pyrochlore spin ice U(1) QSL is described by the emergent compact 
U(1) lattice gauge theory with deconfined and fractionalized 
excitations~\cite{Hermele04,Savary12}. There are three elementary excitations, 
namely, spinon, ``magnetic monopole'', and gauge photon in this U(1) QSL. 
Here the nomenclature for the excitations follows from the original work 
by Hermele, Fisher and Balents~\cite{Hermele04} 
(see Table.~\ref{tab1}). To confirm the realization of the U(1) QSL, 
one would need at least observe one such emergent and exotic excitation. 
Inelastic neutron scattering, that is a spectroscopic measurement, 
is likely to provide much richer information than any other experimental 
probes for the pyrochlore spin ice systems~\cite{Ross11}. It is thus of 
great importance to understand how the neutron moments are coupled to 
the microscopic degrees of freedom and how the inelastic neutron 
scattering (INS) results are related to the emergent and exotic 
properties of the pyrochlore ice U(1) QSL. It is this purpose that 
motivates our work in this paper. 

\begin{table}[b]
\begin{tabular}{p{4cm} p{4cm}}
\hline\hline
 Excitations (notation 1) & Excitations (notation 2)
\\
Spinon & Magnetic monopole 
\\
``Magnetic monopole'' & Electric monopole
\\
Gauge photon & Gauge photon \\
\hline \hline 
\end{tabular}
\caption{Two different but equivalent notations for the excitations 
in the pyrochlore ice U(1) QSL. The notation 1 was introduced 
in Ref.~\onlinecite{Hermele04} and is adopted in this paper.
The notation 2 can be found in Ref.~\onlinecite{PhysRevX.6.011034}, 
and the magnetic monopole in this notation has a classical analogue 
that is a defect tetrahedron with either ``3-in 1-out'' or ``1-in 3-out''
spin configurations~\cite{Castelnovo12008}.}
\label{tab1}
\end{table}

We mainly deal with the non-Kramers doublets in most parts of this paper.
The non-Kramers doublets on the pyrochlore system have been discussed
by several previous works. In particular, the generic spin model 
was introduced and studied in Refs.~\onlinecite{PhysRevLett.105.047201,PhysRevB.83.094411,Sungbin2012},
and more recently, the random strain effect was discussed for Pr$^{3+}$ ions 
in Pr$_2$Zr$_2$O$_7$ in Refs.~\onlinecite{PhysRevLett.118.087203,PhysRevLett.118.107206}. 
In Ref.~\onlinecite{PhysRevB.94.205107}, we have pointed out the magnetic transition 
out of U(1) QSL should be a confinement transition by a simple symmetry analysis. 
For a non-Kramers doublet~\cite{PhysRevLett.105.047201,PhysRevB.83.094411} 
that is described by a pseudospin-1/2 operator $\boldsymbol{S}$, 
the time reversal symmetry, $\mathcal{T}$, acts rather 
peculiarly such that~\cite{Sungbin2012,PhysRevB.94.205107}, 
\begin{eqnarray}
\mathcal{T}: && \quad\quad  S^{x,y} \rightarrow  S^{x,y}, 
\quad 
S^z \rightarrow - S^z. 
\end{eqnarray}
This property means the neutron moments would merely pick up the $S^z$ 
component and naturally measure the $S^z$ correlation. By examining the 
connection with the emergent variables such as gauge fields and 
matter fields, we point out that, the $S^z$ correlation should detect
the gauge photons as well as the ``magnetic monopoles''. 
The ``magnetic monopole'' is the topological defect of the emergent 
vector gauge potential in the compact U(1) quantum electrodynamics
and has no classical analogue. Even though the spinon and the    
``magnetic monopole'' can be interchanged by the electromagnetic 
duality of the lattice gauge theory, the ``magnetic monopole'' might
be more close in spirit to the {\sl Dirac's magnetic monopole}~\cite{Dirac}
from the original definition and theory of the pyrochlore U(1) QSL~\cite{Hermele04}. 
The existence of the ``magnetic monopole'' is one of the key properties 
of the compact U(1) lattice gauge theory~\cite{Fradkinbook} and 
the pyrochlore ice U(1) QSL~\cite{Hermele04}, and it is of great 
importance to demonstrate that the ``magnetic monopole'' is 
a {\sl real physical entity} rather than any artificial or 
fictitious excitation. 

So far, there were only limited studies of ``monopole'' physics in the U(1)
QSL of the pyrochlore ice context~\cite{Hermele04,PhysRevB.94.205107,PhysRevB.95.134439}. 
We here realize that the ``magnetic monopole'' could manifest itself as 
the ``monopole'' continuum in the INS result on the non-Kramers doublet 
pyrochlore spin ice systems. Our renewed understanding of the INS 
measurement for non-Kramers doublets is further extended to the 
Kramers doublets and the quantum Monte carlo 
simulation, and henceforth provides a new insight for the experimental 
observation and the numerical simulation. Moreover, the ``magnetic monopole'' 
experiences a background $\pi$ flux as the ``magnetic monopole'' hops 
around the perimeter on the elementary plaquette of the dual diamond lattice. 
We then point out that the background $\pi$ flux immediately modulates 
the spectral structure of the ``monopole'' continuum by enhancing the  
spectral periodicity. This is an unique experimental signature for the 
``monopole'' continuum in the INS measurement. More generally, this 
is an example of translation symmetry fractionalization in topologically 
ordered phases~\cite{PhysRevB.90.121102,Hermele04,WenPSG}.
Combining with the prior work on the translation symmetry 
fractionalization of the spinons~\cite{GangChen2017}, 
we establish a general classification for the pyrochlore ice U(1) 
QSLs based on the translation symmetry and list their relevant 
spectral properties. 

The following part of the paper is organized as follows. In Sec.~\ref{sec2}, 
we introduce the microscopic model for the non-Kramers doublets, and 
explain the application of several effective models. In Sec.~\ref{sec3}, 
we point out the presence of the ``monopole'' dynamics
in the spin correlation function from the INS measurements. 
In Sec.~\ref{sec4}, we establish the spectral structure of the 
``monopole'' continuum. In Sec.~\ref{sec5}, we carry out the ``monopole'' 
mean field theory and explicitly compute the ``monopole'' dynamics. 
Finally in Sec.~\ref{sec6}, we give a {\sl broad} discussion about the 
spectral properties of non-Kramers doublet and Kramers doublet spin ice materials
and present a classification of the U(1) QSLs based on the translation symmetry 
fractionalization patterns of the ``magnetic monopoles'' and the spinons.

\section{Model for non-Kramers doublets and the low-energy field theory}
\label{sec2}

Due to the peculiar property of the non-Kramers doublets under the time reversal 
symmetry, the generic spin model, that describes the interaction between these 
doublets on the pyrochlore lattice, is actually simpler than the usual Kramers 
doublets and is given by~\cite{PhysRevLett.105.047201,PhysRevB.83.094411,Sungbin2012}
\begin{eqnarray}
H &=& \sum_{\langle ij \rangle} 
{J_{zz}^{} S^z_i S^z_j 
        - J_{\pm}^{} (S^+_i S^-_j + h.c.)}
        \nonumber \\
&+&  {J_{\pm\pm}^{} (\gamma_{ij}^{} S^+_i S^+_j + h.c.)}
+ {\text{dipolar interaction}},
\end{eqnarray}
where ${S^{\pm}_i \equiv S^{x}_i \pm i S^y_i}$ and $\gamma_{ij}$ is the 
bond-dependent phase variable that arises from the spin-orbit-entangled 
nature of the non-Kramers doublet. The dipolar interaction includes the 
further neighbor interactions between the $S^z$ components since only 
$S^z$ is time reversally odd and contributes to the dipole moment. 
It has been shown in Ref.~\onlinecite{Sungbin2012} that, in the perturbative 
Ising limit with ${|{J_{\pm}}| \ll J_{zz}}$ and ${|J_{\pm\pm}| \ll J_{zz}}$, 
the system realizes the U(1) QSL. Moreover, it was demonstrated that 
the U(1) QSL is more robust on the frustrated side~\cite{Sungbin2012}
with ${J_{\pm} < 0}$ and along the axis of $J_{\pm\pm}$. 

Throughout the paper, we deliver our theory through the non-Kramers doublet system.
Only in the Sec.~\ref{sec6}, we extend our theory to the Kramers doublet system.

\subsection{Effective theories}

Our purpose is not to understand the energetics of the relevant
microscopic spin model. We assume that the U(1) QSL has been realized
in the system and try to understand its manifestation in the physical 
observables. For the U(1) QSL, we can then start from the
ring exchange model that is obtained from the perturbative treatment 
of the $J_{\pm}$ and $J_{\pm\pm}$ interactions in the Ising limit. 
With the mapping ${S^z_i = E_{{\boldsymbol r}{\boldsymbol r}'} 
+ \frac{1}{2}}, S^{\pm}_i = e^{\pm i A_{{\boldsymbol r} {\boldsymbol r}'}}$ 
and ${[E_{{\boldsymbol r}{\boldsymbol r}'}, A_{{\boldsymbol r} {\boldsymbol r}'}]=i}$, 
one obtains the U(1) lattice gauge theory on the diamond lattice formed by the 
tetrahedral centers of the pyrochlore lattice~\cite{Hermele04,Savary12}. 
In this lattice gauge theory,
the spinon excitations that violate the ice rule have been traced out 
in the perturbative treatment, and thus, the effective model 
captures the physics below the spinon gap. The lattice gauge theory 
Hamiltonian is given as~\cite{Hermele04}
\begin{eqnarray}
H_{\text{LGT}} = - K \sum_{\hexagon} \cos (curl A) 
                 + \sum_{\langle {\boldsymbol r}{\boldsymbol r}' \rangle } 
                       \frac{U}{2} (E_{{\boldsymbol r}{\boldsymbol r}'} 
                       - \frac{\eta_{\boldsymbol r}}{2} )^2, 
\end{eqnarray}
where ``${\boldsymbol r}, {\boldsymbol r}'$'' stand for the diamond lattice 
sites, ${\eta_{\boldsymbol r} = \pm 1}$ for the two sublattices of the 
diamond lattice, and 
$E_{{\boldsymbol r}{\boldsymbol r}'} = -E_{{\boldsymbol r}'{\boldsymbol r}},
{A_{{\boldsymbol r}{\boldsymbol r}'} = -A_{{\boldsymbol r}'{\boldsymbol r}}}$.
Here, $curl A$ is defined as
\begin{eqnarray}
curl A \equiv{ \sum_{{\boldsymbol r} {\boldsymbol r}' \in 
\hexagon} \!\!\!\!\!\!\!\!\!\!\!\!   \circlearrowleft} \quad A_{{\boldsymbol r} {\boldsymbol r}'},
\end{eqnarray}
and thus corresponds to the magnetic field $B$ through the hexagon center.
The magnetic coupling $K$ is of the order of the ring exchange coupling
in the perturbation theory, and the electric field term is introduced to 
enforce the spin-1/2 Hilbert space. If one focuses on the low-energy and 
long-distance physics, one can further coarsen grain and obtain the 
continuous Maxwell field theory with~\cite{Hermele04}
\begin{eqnarray}
H_{\text{Maxwell}} \simeq \frac{\mathcal{K}}{2} B^2 + \frac{\mathcal{U}}{2} E^2,
\end{eqnarray}
where $\mathcal{K}$ and $\mathcal{U}$ are coarse-grained magnetic and electric couplings.

\begin{figure*}[t]
\centering
\includegraphics[width=0.9\textwidth]{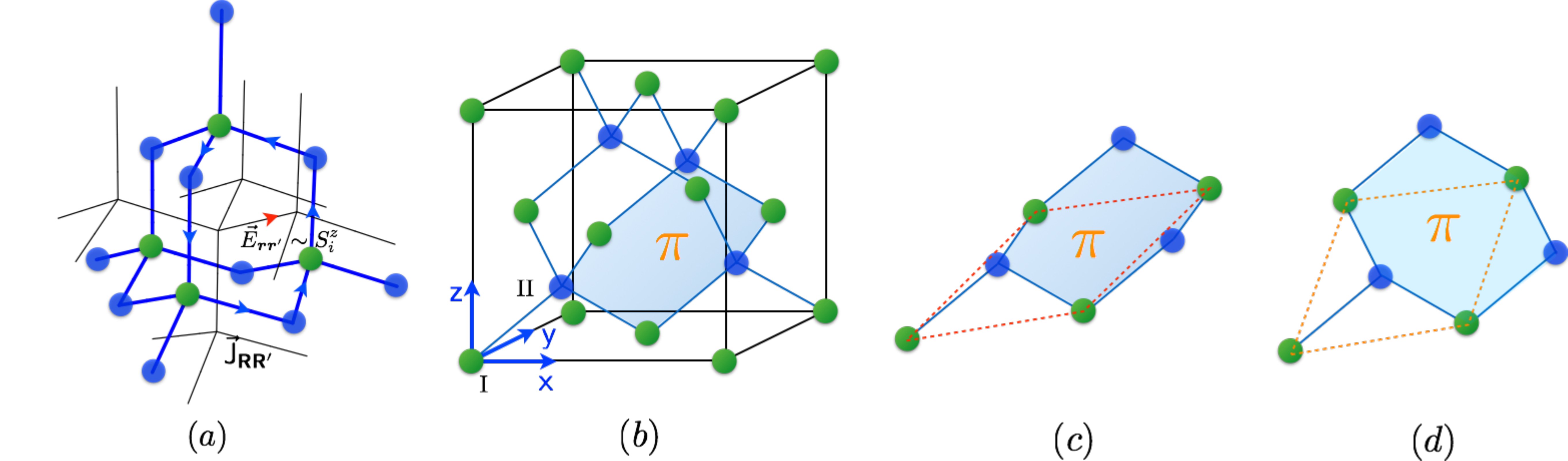}
\caption{(Color online.)
(a) The diamond lattice (in thin line) and its dual diamond lattice (in thick line). 
The physical spin is located in the mid of the link on the diamond lattice. 
The spinons (``monopoles'') hop on the diamond (dual diamond) lattice. 
The colored dots correspond to the tetrahedral centers of the pyrochlore lattice. 
(b) Every buckled hexagon on the dual diamond lattice traps a ``$\pi$'' 
background dual U(1) flux that is experienced by the ``monopole'' hopping. 
``I'' and ``II'' refer to the two sublattices of the dual diamond lattice. 
In (c) and (d), the background flux trapped in the (dashed) parallelogram
is identical to the flux in the (colored) buckled hexagon. 
}
\label{fig1}
\end{figure*}

\subsection{Photon in low-energy theory}

Based on the mapping from the microscopic spin degrees of 
freedom to the emergent field variables in the lattice gauge theory, 
one could establish the connection between the spin correlation function 
with the emergent degrees of freedom. For the non-Kramers doublet, 
the INS measurement would merely pick up the $S^z$ correlator and 
thus measure the correlation function of the emergent electric field. 
It was then shown, within the low-energy Maxwell field theory, 
that the spin correlation corresponds to the electric field 
correlator~\cite{Hermele04,Savary12,PhysRevB.86.075154},
\begin{eqnarray}
\langle {E}^{\mu}_{-{\boldsymbol q},-\omega}  E^{\nu}_{{\boldsymbol q},\omega} \rangle 
\sim
[\delta_{\mu\nu}- \frac{q^{\mu}q^{\nu}}{q^2}] \omega \delta (\omega - v|{\boldsymbol q}|),
\label{eqmax}
\end{eqnarray}
where $v$ is the speed of the photon mode. Apart from the angular dependence, 
the spectral weight of the photon mode is suppressed~\cite{Savary12,PhysRevB.86.075154}
as the energy transfer $\omega \rightarrow 0$.

\section{The loop current of ``magnetic monopoles''}
\label{sec3}

The well-known result of the photon modes in the INS measurement 
was obtained by considering the low-energy field theory that describes 
the long-distance quantum fluctuation within the spin ice manifold. 
The actual spin dynamics, that is captured by the $S^z$ 
correlation in the INS measurement, operates in a broad energy scale 
up to the exchange energy and certainly contains more information 
than just the photon mode from the low-energy Maxwell field 
theory. What is the other information hidden behind? 
To address this question, we have to leave the low-energy Maxwell 
field theory and include the gapped matters into our consideration. 

The gapped matters are spinons and ``magnetic monopoles''. 
The spinons are sources and sinks of the emergent $E$ field
and live on the diamond lattice sites or the tetrahedral centers. 
These spinon are excitations out of the spin ice manifold
and are created by the $S^{x}$ or $S^y$ operator. For the 
non-Kramers' doublet systems, the neutron scattering does not 
allow such spin-flipping processes. So we turn to the 
``magnetic monopoles''. The ``magnetic monopole'' is the 
source or the sink of the emergent $B$ field and is 
the excitation within the spin ice manifold. 
Since the ``magnetic monopole'' is located on the dual 
diamond lattice site (see Fig.~\ref{fig1}), to make the ``magnetic monopole''
explicit, one needs to do a duality transformation on the 
lattice gauge Hamiltonian $H_{\text{LGT}}$~\cite{Hermele04,Bergman2006,PhysRevB.94.205107}. 
This standard procedure~\cite{Hermele04,Bergman2006,PhysRevB.94.205107} 
yields the following dual theory
\begin{eqnarray}
H_{\text{dual}} & = & -t \sum_{\langle \boldsymbol{\mathsf R} \boldsymbol{\mathsf R}' \rangle} 
e^{-i 2\pi \alpha_{\boldsymbol{\mathsf R}\boldsymbol{\mathsf R}'}}
\Phi^{\dagger}_{\boldsymbol{\mathsf R}} \Phi^{\phantom\dagger }_{\boldsymbol{\mathsf R}'} 
-\mu\sum_{\boldsymbol{\mathsf R}} 
\Phi^{\dagger}_{\boldsymbol{\mathsf R}} \Phi^{\phantom\dagger}_{\boldsymbol{\mathsf R}} 
\nonumber \\
&& + \frac{U}{2} \sum_{\hexagon^{\ast}}  (curl \alpha - \frac{\eta_{\boldsymbol r}}{2})^2 
- K \sum_{\langle \boldsymbol{\mathsf R}\boldsymbol{\mathsf R}'\rangle } 
\cos B_{\boldsymbol{\mathsf R}\boldsymbol{\mathsf R}'}^{} 
\nonumber \\
&& + \cdots,
\end{eqnarray}
where ${\Phi}^{\dagger}_{\boldsymbol{\mathsf R}}$ 
(${\Phi}^{}_{\boldsymbol{\mathsf R}}$) creates (annihilates) 
the ``magnetic monopole'' at the dual diamond lattice site 
${\boldsymbol{\mathsf R}}$, ``$\hexagon^{\ast}$'' is the hexagon 
on the dual diamond lattice, ``$t$'' is the ``monopole'' hopping,
and ``$\cdots$'' refers to the ``monopole'' 
interaction. Here $\alpha$ is the dual U(1) gauge field that lives
on the links of the dual diamond lattice, and $curl \, \alpha$ is 
defined as
\begin{eqnarray}
curl \alpha \equiv { \sum_{{\boldsymbol {\mathsf R}} {\boldsymbol {\mathsf R}}' \in 
\hexagon^{\ast}}  \!\! \!\!\! \!\! \!\! \!\! \!\! \circlearrowleft}
\quad
 \alpha_{{\boldsymbol {\mathsf R}} {\boldsymbol{\mathsf R}}'}^{}
\end{eqnarray}
and is simply the electric field going through the center of the hexagon plaquette
on the dual diamond lattice. This dual model describes the coupling between the 
``magnetic monopoles'' and the fluctuating dual U(1) gauge fields, 
and is the starting point to explore the dynamics of the ``magnetic monopoles''. 
For our purpose to capture the generic {\sl spectral structure} of 
the ``monopole'' dynamics, we here keep only the nearest-neighbor 
``monopole'' hopping. 

Since the neutron picks up the $S^z$ component for non-Kramers doublets, 
we want to find what kind of ``monopole'' operators in the dual theory 
correspond to the $S^z$ component. Since this is a gauge theory, only 
gauge invariant quantity is physical according to Elitzur's theorem~\cite{PhysRevD.12.3978}. 
It has been shown from the Maxwell's equations in the early 
studies of critical theories for the ``magnetic monopole'' 
condensation transition~\cite{Bergman2006,PhysRevB.94.205107,PhysRevB.71.125102}, 
that the ``magnetic monopole'' current on a closed hexagon 
loop of the dual diamond lattice induces the electric field 
through the center of the loop (see Fig.~\ref{fig1}), {\sl i.e.}
\begin{eqnarray}
{\sum_{{{\boldsymbol{\mathsf R}}{\boldsymbol{\mathsf R}}' } \in \hexagon^{\ast}} \!\!\!\!
\!\!\!\!\!\!\!\!\!
 \circlearrowleft}
\, \,\, {J}_{{\boldsymbol{\mathsf R}}{\boldsymbol{\mathsf R}}' } = E \sim S^z ,
\end{eqnarray}
where ${J}_{{\boldsymbol{\mathsf R}}{\boldsymbol{\mathsf R}}'}$ 
is the ``monopole'' current between the nearest neighbors with
\begin{eqnarray}
{J}_{{\boldsymbol{\mathsf R}}{\boldsymbol{\mathsf R}}' }^{}
\equiv i [ {\Phi}^{\dagger}_{\boldsymbol{\mathsf R}} 
{\Phi}^{\phantom\dagger}_{\boldsymbol{\mathsf R}'} 
e^{- i 2\pi \alpha_{{\boldsymbol{\mathsf R}}{\boldsymbol{\mathsf R}}' }}
- h.c. ].
\end{eqnarray}
How do we understand the above relation? First, we emphasize that 
this relation is applicable beyond the early studies of identifying   
the proximate {\sl static} $S^z$ Ising order through the ``monopole'' 
condensation, and holds even for the {\sl dynamical property 
inside the U(1) QSL phase}. Second, there is no contradiction between 
this relation with Eq.~(\ref{eqmax}) that is a coarse-grained low-energy 
and long-distance result. This relation here includes the short distance
and finite energy dynamics of the ``magnetic monopoles''. 
From this relation, we conclude that the $S^z$ correlation 
contains the contribution of the ``monopole'' current correlator.  

The above analysis does not provide the information about the spectral 
weight of the ``monopole'' continuum in the $S^z$ correlation. 
It was pointed out that increasing further neighbor $S^z$-$S^z$ 
interaction could drive a quantum phase transition from the U(1) QSL 
to the Ising order via the ``monopole'' condensation~\cite{PhysRevB.94.205107}. 
We thus think that the systems with extended $S^z$ coupling may have 
more visible ``monopole'' continuum in the INS result. 

\section{The spectral structure of the ``monopole'' continuum}
\label{sec4}

We realize that the physical spin operator, $S^z$, creates one 
``monopole''-``anti-monopole'' pair. The dynamic spin structure 
factor of the non-Kramers doublet would contain a broad ``monopole'' 
continuum due to this ``fractionalization'' of the spin into the 
two ``monopoles''. Here we are interested in the generic and unique 
spectral structure rather than some specific details that can be 
used to uniquely identify the ``monopole'' continuum in the INS 
results.  

The ``magnetic monopole'' hops on the dual diamond lattice and   
experiences the dual U(1) gauge flux. The background gauge flux 
thus {\sl modulates} the ``monopole'' dynamics. Due to the electric 
field offset, ${\eta_{\boldsymbol r}/2}$, that originates fundamentally 
from the effective spin-1/2 nature of the local moment, there 
exists a background gauge flux on each hexagon plaquette 
of the dual diamond lattice with~\cite{PhysRevB.94.205107}
\begin{eqnarray}
2\pi \langle curl \, \alpha \rangle = {\pi \eta_{\boldsymbol r} \equiv \pi}
\quad (\text{mod}\, 2\pi).
\end{eqnarray}

To see the effect of the background dual gauge flux, we introduce 
the translation operator for the ``magnetic monopole'', $T^{m}_{\mu}$,
that translates the ``monopole'' by a basis lattice vector     
${\boldsymbol {\mathsf a}}_{\mu}$ of the dual diamond lattice, 
where ${\mu=1,2,3}$, and 
${{\boldsymbol {\mathsf a}}_{1}= \frac{1}{2}(011)}, 
{{\boldsymbol{\mathsf a}}_{2}=\frac{1}{2}(101)},
{{\boldsymbol{\mathsf a}}_{3}=\frac{1}{2}(110)}$. 
We use the cubic coordinate system and set the lattice 
constant to unity throughout the paper. 
As the ``magnetic monopole'' hops successively 
through the parallelogram defined by 
$T^m_{\mu} T^m_{\nu} (T^m_{\mu})^{-1} (T^m_{\nu})^{-1}$ with 
${\mu \neq \nu}$, the ``monopole'' experiences an identical 
Aharonov-Bohm flux as the background flux trapped in the hexagon 
plaquette of the dual diamond lattice (see Fig.~\ref{fig1}). 
This is because of the lattice geometry of the diamond lattice. 
Thus, we have the following algebraic relation
\begin{eqnarray}
T^m_{\mu} T^m_{\nu} (T^m_{\mu})^{-1} (T^m_{\nu})^{-1} = e^{i \pi} = -1.
\label{anticomm}
\end{eqnarray}
This algebraic relation means the lattice translation symmetry is realized 
{\sl projectively} for the ``magnetic monopoles''.  
The translation symmetry fractionalization for the ``magnetic monopole'' 
is intimately connected to the spectral periodicity of the ``monopole continuum''
~\cite{PhysRevB.90.121102,WenPSG,Wen2002175}. 

To demonstrate the enhanced spectral periodicity of the ``monopole'' 
continuum, we introduce a 2-``monopole'' scattering state
${|\text{A} \rangle \equiv |{\boldsymbol {\mathsf q}}_{\text A}; 
{\mathsf z}_{\text A} \rangle}$, where ${\boldsymbol {\mathsf q}}_{\text A}$ 
is the total crystal momentum of this state and ${\mathsf z}_{\text A}$ 
represents the remaining quantum number that specifies the state~\cite{PhysRevB.90.121102}. 
The translation symmetry fractionalization acts on the individual ``monopole'', 
such that 
\begin{eqnarray}
T_{\mu} |\text{A} \rangle \equiv T_{\mu}^m (1) T_{\mu}^m (2) |\text{A} \rangle  ,
\end{eqnarray}
where $T_{\mu}$ is the translation operator for the system, and 
``1'' and ``2'' refer to the two ``monopoles'' of this state. By translating
one ``monopole'' by the basis lattice vector ${\boldsymbol{\mathsf a}}_{\mu}$, 
we obtain another three 2-``monopole'' scattering states,
\begin{eqnarray}
|\text{B} \rangle & = &  T^m_1 (1) |\text{A} \rangle,\\
|\text{C} \rangle & = &  T^m_2 (1) |\text{A} \rangle,\\
|\text{D} \rangle & = &  T^m_3 (1) |\text{A} \rangle .  
\end{eqnarray}

It is ready to compare the translation eigenvalues of these
four states by making use of Eq.~(\ref{anticomm}) and obtain the following
relations for the crystal momentum of these states, 
\begin{eqnarray}
{\boldsymbol {\mathsf q}}_{\text B} & = & {\boldsymbol {\mathsf q}}_{\text A} + 2\pi (100) , \\
{\boldsymbol {\mathsf q}}_{\text C} & = & {\boldsymbol {\mathsf q}}_{\text A} + 2\pi (010) , \\
{\boldsymbol {\mathsf q}}_{\text D} & = & {\boldsymbol {\mathsf q}}_{\text A} + 2\pi (001). 
\end{eqnarray}
Since these scattering states have the same energy, we thus conclude that 
the ``monopole continuum'' of the two ``monopole'' excitations  
have the following enlarged spectral periodicity such that
\begin{eqnarray}
{\mathsf L}_{m} ({\boldsymbol {\mathsf q}}  ) & = & 
{\mathsf L}_m ({\boldsymbol {\mathsf q}} + 2\pi (100)  )  
\nonumber \\
& = & {\mathsf L}_m ({\boldsymbol {\mathsf q}} + 2\pi (010)  ) 
\nonumber \\
& = & {\mathsf L}_m ({\boldsymbol {\mathsf q}} + 2\pi (001) ) ,
\end{eqnarray}
where ${\mathsf L}_{m} ({\boldsymbol {\mathsf q}} )$ is the lower
excitation edge 
of the ``monopole'' continuum for a given momentum ${\boldsymbol {\mathsf q}}$
because there is a finite energy cost to excite two ``monopoles''.
This enhanced spectral periodicity also appears in the upper excitation 
edges of the ``monopole'' continuum.
There is no symmetry breaking nor any static magnetic order in the system,
but the spectral periodicity is enhanced. The spectrum is 
invariant if one translates the spectrum by $2\pi (100)$, $2\pi (010)$,
or $2\pi (001)$. This is very different from the conventional case 
where the spectral periodicity is given by
the reciprocal lattice vectors, $2\pi (\bar{1}11)$,
$2\pi (1\bar{1}1)$ and $2\pi (11\bar{1})$, for the FCC bravais lattice. 
Therefore, the spectral periodicity enhancement 
with a fold Brillouin zone is a strong indication of 
the fractionalization in the system.

\section{The ``monopole'' mean-field theory and the continuum}
\label{sec5}

To explicitly compute the ``monopole'' dynamics and demonstrate the 
spectral periodicity enhancement, we carry out the mean-field 
approximation for the ``monopole''-gauge coupling. To capture
the $\pi$ background flux, we set the dual gauge potential as~\cite{Sungbin2012,PhysRevB.94.205107} 
\begin{eqnarray}
2\pi \langle \alpha^{}_{{\boldsymbol{\mathsf R}},{\boldsymbol{\mathsf R}} 
+ {\boldsymbol{\mathsf e}}_{\mu}} \rangle 
= {\mathsf{\xi}}_{\mu} ({\boldsymbol{\mathsf Q}} \cdot {\boldsymbol{\mathsf R}}),
\end{eqnarray}
where ${\boldsymbol {\mathsf R}} \in$ I sublattice of the dual diamond lattice, and
${{\boldsymbol{\mathsf R}} + {\boldsymbol{\mathsf e}}_{\mu} }\in$ II sublattice of 
the dual diamond lattice with ${\boldsymbol{\mathsf e}}_{\mu}$ (${\mu=0,1,2,3}$) 
the nearest-neighbor vectors connecting two sublattices. 
Here $\boldsymbol{\mathsf{e}}_{\mathsf 0} = \frac{1}{4}(111),
\boldsymbol{\mathsf e}_{\mathsf 1} = \frac{1}{4}(1\bar{1}\bar{1}),
\boldsymbol{\mathsf e}_{\mathsf 2} = \frac{1}{4}(\bar{1}1\bar{1}),
\boldsymbol{\mathsf e}_{\mathsf 3} = \frac{1}{4}(\bar{1}\bar{1}1)$,  
${(\xi_{\mathsf 0},\xi_{\mathsf 1},\xi_{\mathsf 2},\xi_{\mathsf 3} ) 
=(0,1,1,0)}$ and ${\boldsymbol{\mathsf Q}} = 2\pi(100)$. 

Under this above gauge fixing, we have the ``monopole'' mean-field Hamiltonian,
\begin{eqnarray}
{H_{\text{MFT}} = 
-t \sum_{\langle \boldsymbol{\mathsf R} \boldsymbol{\mathsf R}' \rangle} 
e^{-i 2\pi \langle \alpha_{\boldsymbol{\mathsf R}\boldsymbol{\mathsf R}'} \rangle }
\Phi^{\dagger}_{\boldsymbol{\mathsf R}} \Phi^{\phantom\dagger }_{\boldsymbol{\mathsf R}'} 
-\mu\sum_{\boldsymbol{\mathsf R}} 
\Phi^{\dagger}_{\boldsymbol{\mathsf R}} \Phi^{\phantom\dagger}_{\boldsymbol{\mathsf R}} } ,
\end{eqnarray}
where the ``monopole'' spectrum is found to be
\begin{eqnarray}
\Omega^+_{\pm} ({\boldsymbol {\mathsf q}}) &=& +t [4\pm 2 (3+ 
{\mathsf C}_{\mathsf x} {\mathsf C}_{\mathsf y} - 
{\mathsf C}_{\mathsf x} {\mathsf C}_{\mathsf z} + 
{\mathsf C}_{\mathsf y} {\mathsf C}_{\mathsf z})^{\frac{1}{2}}]^{\frac{1}{2}}- \mu ,
\nonumber 
\\
\Omega^-_{\pm} ({\boldsymbol {\mathsf q}}) &=& -t [4\pm 2 (3+ 
{\mathsf C}_{\mathsf x} {\mathsf C}_{\mathsf y} - 
{\mathsf C}_{\mathsf x} {\mathsf C}_{\mathsf z} + 
{\mathsf C}_{\mathsf y} {\mathsf C}_{\mathsf z})^{\frac{1}{2}}]^{\frac{1}{2}}- \mu ,
\nonumber
\end{eqnarray}
where ${{\mathsf C}_{\mu} = \cos {\mathsf q}_{\mu} }$ (${\mu = {\mathsf{x,y,z}}}$). 
There are four ``monopole'' bands: two arise from the two sublattices of 
the dual diamond lattice, and two arise from the gauge fixing that doubles 
the unit cell. 

\begin{figure}
\includegraphics[width=5.8cm]{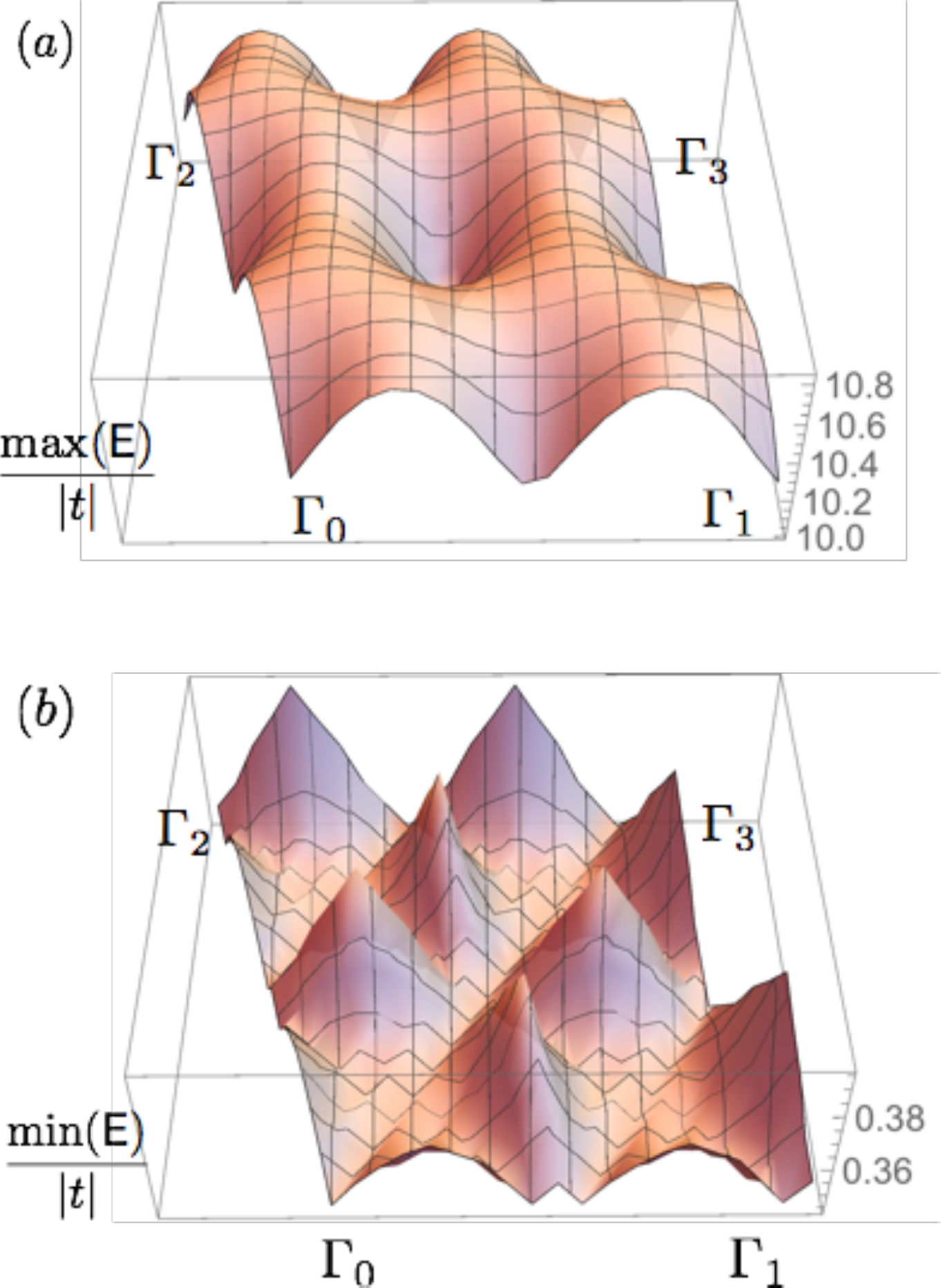}
\caption{(Color online.) (a) The upper excitation edge of the ``monopole'' continuum.
(b) The lower excitation edge of the ``monopole'' continuum. 
For both figures, we set ${\mu = -3t}$, and the $\Gamma$ points
are the Brillouin zone centers. The important information of the plot 
is not the dispersion itself, instead is the enhanced spectral 
periodicity as if the Brillouin zone is folded. Here
${\Gamma_0\Gamma_1} = 2\pi (\bar{1}11)$
and ${\Gamma_0\Gamma_2} = 2\pi(1\bar{1}1)$ are the 
reciprocal lattice vectors. }
\label{fig2}
\end{figure}

As we point out in Sec.~\ref{sec4}, the ``monopole'' continuum 
is contained in the ``monopole'' current correlation. 
Here we are interested in 
the spectral structure of the upper and lower excitation edges
of the ``monopole'' continuum. From the momentum and the energy 
conservation, we have for the two ``monopoles''
\begin{eqnarray}
{\boldsymbol{\mathsf q}} & = & {\boldsymbol{\mathsf q}}_{\mathsf 1} + 
{\boldsymbol{\mathsf q}}_{\mathsf 2} + {\boldsymbol{\mathsf Q}}, \\ 
{\mathsf E} & = & \Omega^{i_1}_{j_1} ({\boldsymbol{\mathsf q}}_{\mathsf 1}) + 
\Omega^{i_2}_{j_2} ({\boldsymbol{\mathsf q}}_{\mathsf 2}),
\end{eqnarray}
where ${\boldsymbol{\mathsf q}}$ and ${\mathsf E}$ are the momentum and energy 
transfer of the neutrons, ${\boldsymbol{\mathsf q}}_{\mathsf 1}$ and 
${\boldsymbol{\mathsf q}}_{\mathsf 2}$ are the crystal momenta of the 
two ``monopoles'', and the offset ${\boldsymbol{\mathsf Q}}$ arises
from the dual gauge link that is present in the ``monopole'' current. 
The minimum (maximum) of the energy ${\mathsf E}$
is obtained when ${i_1 = i_2 = -}$ and ${j_1=j_2=+}$ 
(${i_1 = i_2 = +}$ and ${j_1=j_2=+}$).
In Fig.~\ref{fig2}, we depict the upper and lower excitation edges 
of the ``monopole'' continuum for a specific choice of 
``monopole'' hopping and chemical potential. Clearly, 
the spectral periodicity is enhanced in both plots.  

\section{Discussion}
\label{sec6}

\subsection{Non-Kramers doublets}

We discuss the application of our results to 
 various pyrochlore ice systems. We begin with 
the non-Kramers doublets. The continuous excitations 
have actually been observed from the INS measurements 
on Pr$_2$Zr$_2$O$_7$, Tb$_2$Ti$_2$O$_7$ and Pr$_2$Hf$_2$O$_7$~\cite{Romain2017,PhysRevLett.118.107206,0953-8984-24-5-052201}.
In particular, in the INS result for Pr$_2$Hf$_2$O$_7$~\cite{Romain2017}, 
besides the very low-energy features that seem to resemble the suppressed 
spectral intensity of the photon mode, there exists a broad excitation 
continuum extending to higher energies. 
This continuum may be attributed to the random strain effect that has 
already been suggested to Pr$_2$Zr$_2$O$_7$~\cite{Baker156,PhysRevLett.118.107206,PhysRevLett.118.087203}. 
Nevertheless, the random strain effect was also suggested to create
quantum entanglement and induce U(1) QSL phase in non-Kramers doublet 
systems~\cite{PhysRevLett.118.087203}. Therefore, if the underlying
systems realize the U(1) QSL, according to our theory, 
these mysterious continuous excitations may at least contain the 
contribution from the two-``monopole'' continuum that is predicted
in this work.  

How does one verify the above claim of the ``monopole'' continuum 
in the INS measurement? We here propose a scheme to exclude the presence 
of the spinon continuum in the INS result by conducting a thermal 
transport measurement. Spinons are higher energy excitations, 
and their contribution to thermal conductivity should appear 
at higher temperatures~\cite{Matsuda}. If one observes that 
the energy scale of the continuum in the INS measurement is clearly lower 
than the temperature scale where the spinons contribute to the thermal 
conductivity, one could then conclude the presence of the spinon excitation 
in the thermal conductivity results and the absence of the spinon excitation 
in the continuum of the INS results. The direct measurement would 
be the confirmation of the enhanced spectral periodicity of the
``monopole'' continnum in the momentum space. This may be difficult
as the low-energy photon excitation is also present in the low-energy 
INS data. Thus, the higher energy part of the ``monopole'' continnum may 
provide more useful information. It is certainly very exciting if all 
the three excitations, spinon, ``magnetic monopole'', and gauge photon 
are confirmed by a combination of the INS and the thermal transport 
measurements. 

For the ``monopoles continuum'', probably the most positive side 
in this identification of  ``monopole continuum'' is that 
weak external magnetic field can be used to manipulate the ``monopole''
continuum. With weak magnetic fields, the U(1) QSL will not be destroyed,
and the ``magnetic monopole'' remains to be a valid description of the 
excitation of the system. However, the external magnetic field, that 
only couples linearly to the $S^z$ components, polarizes $S^z$ slightly
and thereby modifies the background dual U(1) gauge flux 
that is experienced by the ``monopole''. As a result, 
the ``monopole'' band would probably develop a Hofstadter band~\cite{PhysRevB.14.2239}, 
and the spectral structure of the ``monopole'' continuum is modified. 
How this ``monopole'' continuum is modulated depends on the orientation 
and the amplitude of the external magnetic fields. The detailed behavior 
of the ``monopole'' continuum in the weak field will be explored in future works.

\begin{table}[t]
\begin{tabular}{lcc}
\hline\hline
Properties & U(1)$_{0,\pi}$ QSL & U(1)$_{\pi,\pi}$ QSL \\
spinon flux   & $0$  & $\pi$        \\
``monopole'' flux & $\pi$  &  $\pi$ \\
spinon continuum & not enhanced & enhanced \\
``monopole'' continuum & enhanced  & enhanced 
\\
\hline\hline
\end{tabular}
\caption{A classification of distinct U(1) QSLs from the symmetry 
classification patterns of the spinons and the ``magnetic monopoles''. 
The first subindex refers to the flux that is experienced by 
the spinon hopping around the hexagon plaquette on the diamond 
lattice (see the second row), while the second subindex refers to 
the flux that is experienced by the ``monopole'' hopping around 
the hexagon plaquette on the dual diamond lattice (see the third row). 
In the table, ``enhanced'' and ``not enhanced'' refer
to the spectral periodicity of the related excitation continuum. }
\label{tab2}
\end{table}

\subsection{Kramers doublets and numerical simulation}

As for the usual Kramers doublets~\cite{Ross11,Savary12,Gingras2014}, 
all the three components of the local moments 
are odd under the time reversal symmetry, and the neutron spin would couple 
to all of them. Therefore, the INS results on the U(1) QSL with the usual Kramers
doublets would also detect the spin flipping events out of the spin ice manifold 
and measure the spinon continuum in addition to the gauge photon and the 
``monopole'' continuum. As we have already pointed out in the previous 
sections, the visibility of the ``monopole'' continuum in the INS data depends
on how much weight of the ``monopole'' continuum, and may vary for different materials.  

If the neutron energy transfer is located within the ``monopole'' 
continuum, the spectral periodicity would experience an enhancement. 
If the neutron energy transfer is located in the spinon continuum, 
the spectral periodicity is enhanced (not enhanced) 
if the spinon experiences a background ${\pi}$ ($0$) 
flux on the diamond lattice~\cite{GangChen2017}. 

The U(1) QSL has been explored by quantum Monte carlo simulation,
and the photon mode was identified in the $S^z$ correlation 
function~\cite{PhysRevLett.115.037202,PhysRevLett.115.077202,PhysRevLett.100.047208}. 
It might be of interest to introduce further $S^z$ interactions to 
possibly enhance and manifest the ``monopole'' continuum
in the $S^z$ correlation~\cite{PhysRevB.94.205107}.

\subsection{A classification of the U(1) QSLs}

Finally, let us remark on the translation symmetry fractionalization patterns 
for the U(1) QSLs. In this work, we have focused on the ``magnetic monopole'' 
excitation and found that the ``magnetic monopole'' experiences a background 
dual U(1) flux on the dual diamond lattice. In the previous work~\cite{GangChen2017}, 
we studied the spectral periodicity
and the translation symmetry fractionalization for the spinon excitation. 
The combination of the ``magnetic monopole'' and the spinon symmetry fractionalization 
patterns results in a classification of the distinct symmetry enriched U(1) 
QSLs in Table~\ref{tab2}. Like the classification scheme that was developed for 
the two-dimensional $\mathbb{Z}_2$ QSLs and applied to the $\mathbb{Z}_2$ 
toric code model~\cite{PhysRevB.87.104406}, 
one could use the result in Table~\ref{tab2} 
to further establish the translation symmetry fractionalization for the (fermionic) 
dyon that is a bound state of the spinon and the ``monopole''.
Our classification not only helps improve the understanding 
of the crystal symmetry fractionalization in the U(1) QSLs, but also 
provides unique and detectable experimental signatures for the U(1) QSLs.

\section{Acknowledgments}

We acknowledege Nic Shannon and Mike Hermele for useful discussion, Chenjie Wang 
for various related and unrelated philosophical conversations, Zhong Wang for a 
comment, and one anonymous referee for a comment that improves this work. We acknowledge 
Michel Gingras for the invitation to the ``International Workshop on Quantum Spin Ice'' 
at Perimeter Institute for Theoretical Physics where this work is carried out and 
finalized. This work is supported by the ministry of science and technology of 
China with the Grant No.2016YFA0301001, the start-up fund for original research 
and the first-class university construction fund of Fudan University, and the 
thousand-youth-talent program of China.

\bibliography{ref}

\end{document}